\documentclass{iau} 
\usepackage{graphicx}

\def\lesssim{\mathrel{\hbox{\rlap{\hbox{\lower4pt\hbox{$\sim$}}}\hbox{$<$}}}}
\def\gtrsim{\mathrel{\hbox{\rlap{\hbox{\lower4pt\hbox{$\sim$}}}\hbox{$>$}}}}
\def \aj {AJ}
\def \mnras {MNRAS}
\def \apj {ApJ}
\def \apjs {ApJS}

\def \aap {A\&A}
\def \aaps {A\&AS}

\def \araa {ARAA}

\def \nar {New Astron. Rev.}

\title[SBF for constraining the chemical enrichment] 
{Surface Brightness Fluctuations for constraining the chemical enrichment of massive galaxies}

\author[Vazdekis et al.]   
{A. Vazdekis$^{1,2}$,
P. Rodr\'{\i}guez-Beltr\'an$^{1,2}$, M. Cervi\~no$^{3}$, M. Montes$^{4}$, I. Mart\'{\i}n-Navarro$^{5,6}$
 \and M.~B. Beasley$^{1,2}$}

\affiliation{$^1$Instituto de Astrof{\'{\i}}sica de Canarias, E-38200
La Laguna, Tenerife, Spain \\ email: {\tt vazdekis@iac.es} 
\\[\affilskip]
$^2$Departamento de Astrof{\'{\i}}sica, Universidad de La Laguna,
E-38205, Tenerife, Spain\\
$^{3}$Centro de Astrobiolog{\'{\i}}a (CSIC/INTA), ESAC Campus, Camino Bajo del Castillo s/n, E-28692 Villanueva de la Ca\~nada, Spain\\
$^{4}$School of Physics, University of New South Wales, Sydney, NSW 2052, Australia\\
$^{5}$Max-Planck Institut f\"ur Astronomie, Konigstuhl 17, D-69117 Heidelberg, Germany\\
$^{6}$University of California Santa Cruz, 1156 High Street, Santa Cruz, CA 95064, USA
}

\pubyear{2020}
\volume{359}  
\jname{Galaxy evolution and feedback across different environments}
\editors{T. Storchi-Bergmann, R. Overzier, W. Forman \& R. Riffel, eds.}
\begin{document}

\maketitle

\begin{abstract}
Based on very deep photometry, Surface Brightness Fluctuations (SBF) have been  traditionally used to determine galaxy distances. We have recently computed SBF spectra of stellar populations at moderately high resolution, which are fully  based on empirical stars. We show that the SBF spectra provide an unprecedented  potential for stellar population studies that, so far, have been tackled on the basis of the mean fluxes. We find that the SBFs are able to unveil  metal-poor stellar components at the one percent level, which are not possible to disentangle with the standard analysis. As these metal-poor components correspond to the first stages of the chemical enrichment, the SBF analysis provides stringent constrains on the quenching epoch.

\end{abstract}

\firstsection 
\section{Introduction}

Stellar populations are characterized by differences in the luminosity
distribution of stars contributing to the flux in a given resolution element. By normalizing the variance, i.e. the second moment, of these fluctuations by the mean flux, i.e. the first moment, in that element we obtain the so called SBF (\cite[Tonry \& Schneider 1988; Tonry, Ajhar \& Luppino 1990]{TS88,Tonry90}). The SBF is an intrinsic property of an SSP, i.e. a simple, single-burst, stellar
population characterized by a single-age and single-metallicity and, therefore, it depends on these parameters. For obtaining accurate fluctuation magnitudes we need very high quality photometry as well as subtracting a smooth galaxy model and it is often applied a Fourier Transform analysis to isolate the intrinsic fluctuations of the stellar populations.

So far the main application of the SBF method it has been its use for obtaining accurate galaxy
distances in the nearby Universe, as the observed flux of the fluctuations
depends on galaxy distance, with more distant galaxies appearing smoother. The SBFs provided distances to Virgo and Fornax with a precision of $2\%$
(e.g., \cite[Blakeslee et~al. 2010]{Blakeslee10}). SBFs have also been shown to provide additional constrains to relevant stellar population parameters (\cite[Liu, Charlot \& Graham 2000; Blakeslee, Vazdekis \& Ajhar 2001]{Liu00,Blakeslee01}). In fact, by confronting theoretical SBF model predictions with the observational fluctuations it is possible to break the  age/metallicity degeneracy (\cite[Worthey 1994; Cantiello et~al. 2003]{Worthey94,Cantiello03}). However such applications are very scarce in the literature, mostly due to the lack of SBF determinations in more than just a single band for a given object.

As it happens with the mean fluxes, the SBF method can be also used in spectroscopic studies, with increased abilities to break main stellar population degeneracies through key spectral features. Low resolution theoretical SBF spectra can be derived from the predictions of \cite[Buzzoni (1993)]{Buzz93} or at high resolution \cite[Gonzalez-Delgado et~al. (2005)]{GDetal05} as shown in \cite[Cervi\~no (2013)]{Cer13}. These models are based on fully theoretical stellar spectral libraries. Recently were presented model SBF spectra at moderately high resolution (\cite[Mitzkus et~al. 2018; Vazdekis et~al. 2020]{Mitzkus18,Vazdekis20}), based on empirical stellar libraries. Very recently it has been presented the first observational SBF spectrum of a nearby S0 galaxy \cite[Mitzkus et~al. (2018)]{Mitzkus18}, using data from Multi Unit Spectroscopic Explorer (MUSE) Integral Field Spectrograph instrument. Although promising, much work is required to propose a robust treatment of the data to obtain the observational SBF spectra as well as to define an optimal methodology to extract the information contained in these spectra (\cite[Vazdekis et~al. 2020]{Vazdekis20}). Here we employ fluctuation colour-colour plots from our recently computed E-MILES (Extended - MILes de EStrellas) SBF model spectra to study the first stages of the evolution of Early-Type Galaxies (ETGs). This allows us to constrain the very first stages of their chemical evolution.

\section{Models}

Here we employ the E-MILES SBF model spectra\footnote{Models can be downloaded from the MILES website: http://miles.iac.es/} presented in \cite[Vazdekis et~al. 2020]{Vazdekis20}. Briefly, these models combine the isochrones of \cite[Girardi et~al. (2000)]{Padova00} and \cite[Pietrinferni et~al. (2004)]{Pietrinferni04} with fully empirical stellar spectral libraries. These libraries include the Hubble Space Telescope based Next Generation Stellar Library (\cite[NGSL, Gregg et~al. 2006]{NGSL}), Medium-resolution Isaac Newton telescope Library of Empirical Spectra (\cite[MILES, S{\'a}nchez-Bl{\'a}zquez et~al.2006]{MILESI}), Indo-US (\cite[Valdes et~al. 2004]{Valdes04}), Calcium Triplet (\cite[CaT, Cenarro et~al. 2001]{CATI}) and Infra-Red Telescope Facility library (\cite[IRTF, Cushing, Raynier \& Vacca 2005; Raynier, Cushing \& Vacca 2009]{IRTFI,IRTFII}). The models are computed for a suite of Initial Mass Function (IMF) shapes and slopes (\cite[Kroupa 2001; Chabrier 2001; Vazdekis et~al. 1996]{Kroupa01,Chabrier,Vazdekis06}. The various spectral ranges covered by these models are joined as described in \cite[Vazdekis et~al. (2016)]{Vazdekis16} to build-up both the mean and the SBF extended E-MILES spectra at moderately high resolution.

Working with SBFs has some peculiarities to take into account as extensively described in \cite[Vazdekis et~al. (2020)]{Vazdekis20}. This concerns the modelling of more complex stellar populations and obtaining spectroscopic SBF magnitudes. In brief, due to the properties of the variance, we cannot combine the SBF spectra of the SSPs that contribute to a composite stellar population to obtain its SBF spectrum. The two, the SSP variance and the SSP mean spectra need to be combined separately and, only then, it is possible to divide them to obtain the composite SBF spectrum. Another important peculiarity to take into account concerns the obtention of SBF magnitudes from the SBF spectra. In this case we cannot simply convolve the SBF spectrum with the response of the desired filter, as there are correlations among the resolution elements. Several options were investigated in \cite[Vazdekis et~al. (2020)]{Vazdekis20} to conclude that, under the very conservative hypothesis of full correlation among pixels, the spectroscopic SBF magnitude can be derived as   

\begin{equation}
  \label{eq:SBF_band2}
  \bar{m}(\mathrm{spec}) = 2 \times m_{\sqrt{F^\mathrm{var}_\lambda}} - m_{F^\mathrm{mean}_\lambda}
\end{equation}
  
  \noindent where $\bar{m}$ stands for the SBF magnitude,
  and $m_{\sqrt{F^\mathrm{var}_\lambda}}$ and $m_{F^\mathrm{mean}_\lambda}$ are
  the magnitudes obtained by convolving the filter response with the square root of the SSP variance and the mean SSP, respectively.

\section{Results}

\begin{figure}[b]
\begin{center}
 \includegraphics[width=5.4in]{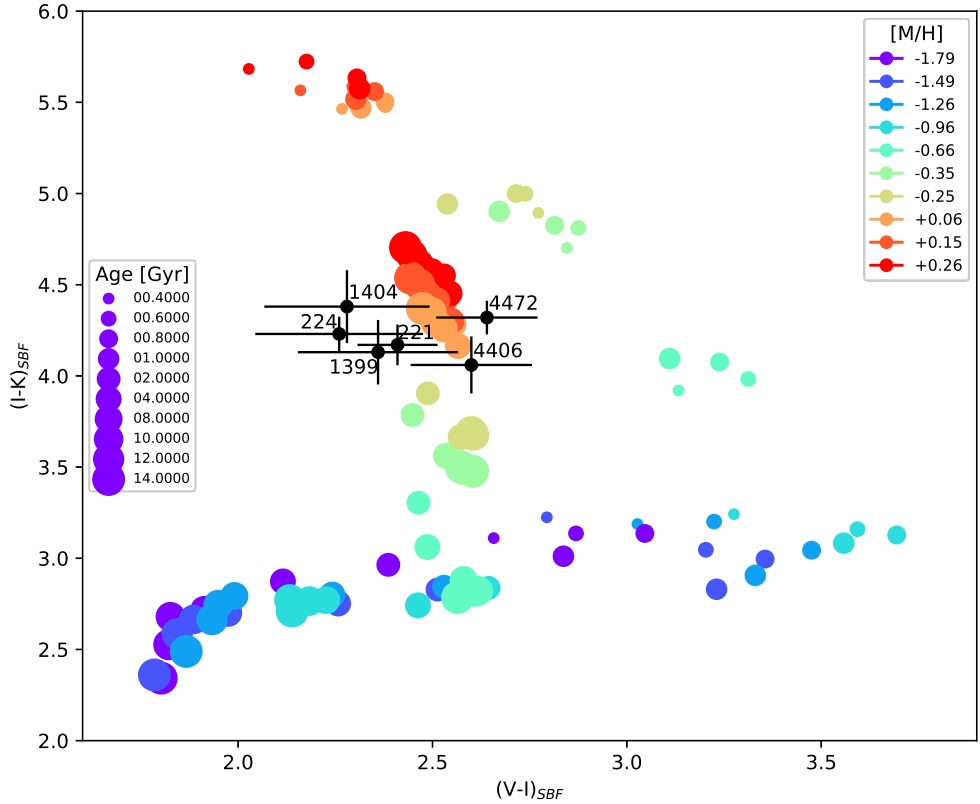} 
 \caption{$V-I$ vs. $I-K$ fluctuation colour-colour diagram. The solid coloured circles represent the SBF colours corresponding to SSPs with varying metallicity [M/H] (from very metal-poor to metal-rich, i.e. from purple to red, respectively, as indicated in the top-right inset) and varying age (increasing from $0.4$ to $14$\,Gyr, with increasing circle size as indicated in the bottom-left inset). The SBF data corresponding to a number of representative ETGs are shown in black solid circles, including their errorbars. Note that a fraction of the galaxies fall in a region that is not matched by the SSPs.  
 }
   \label{fig1}
\end{center}
\end{figure}

According to the mean photometric and spectroscopic properties of massive ETGs these galaxies are found to be old and metal-rich \cite[(Renzini 2006)]{Renzini06}. Fig.~\ref{fig1} illustrates the potential of the SBFs for further constraining the stellar populations properties of these galaxies. A number of representative ETGs from \cite[Cantiello et~al. (2003)]{Cantiello03} are shown in the $V-I$ vs. $I-K$ fluctuation colour-colour diagnostic diagram. We see that a fraction of these ETGs fall in a region that is not occupied by any of the old SSPs of varying metallicities. Only combinations of very metal-rich and very-metal poor SSPs are able to match these galaxies. As shown in \cite[Vazdekis et~al. (2020)]{Vazdekis20} ,mass fractions of $1-5\%$ of stellar populations with metalliciticies $[Fe/H]<-1$ are required on the top of the largely dominating old metal-rich stellar population to be able to match this set of galaxies. 

\section{Discussion}

So far, these very small contributions from very metal-poor components have been elusive to the standard analysis based on the mean fluxes. Note that these contributions are not related to the well known age/metallicity degeneracy affecting the bulk of the population (\cite[Worthey 1994]{Worthey94}). These small components, which have been disentangled by the SBF diagnostic diagram, must be associated to the first stages of galaxy chemical enrichment. In fact, it is remarkable the agreement between these results and the full chemo-evolutionary modelling of \cite[Vazdekis et~al. (1996)]{Vazdekis06}. These models predicted a rapid ($<200$\,Myr) enrichment for the innermost regions of massive ETGs, formed in-situ, leading to residual mass-fractions of very metal-poor stellar populations smaller than $5\%$. This light come from  long-lived low-mass stars formed at $z>2$.

\begin{figure}[b]
\begin{center}
 \includegraphics[width=5.4in]{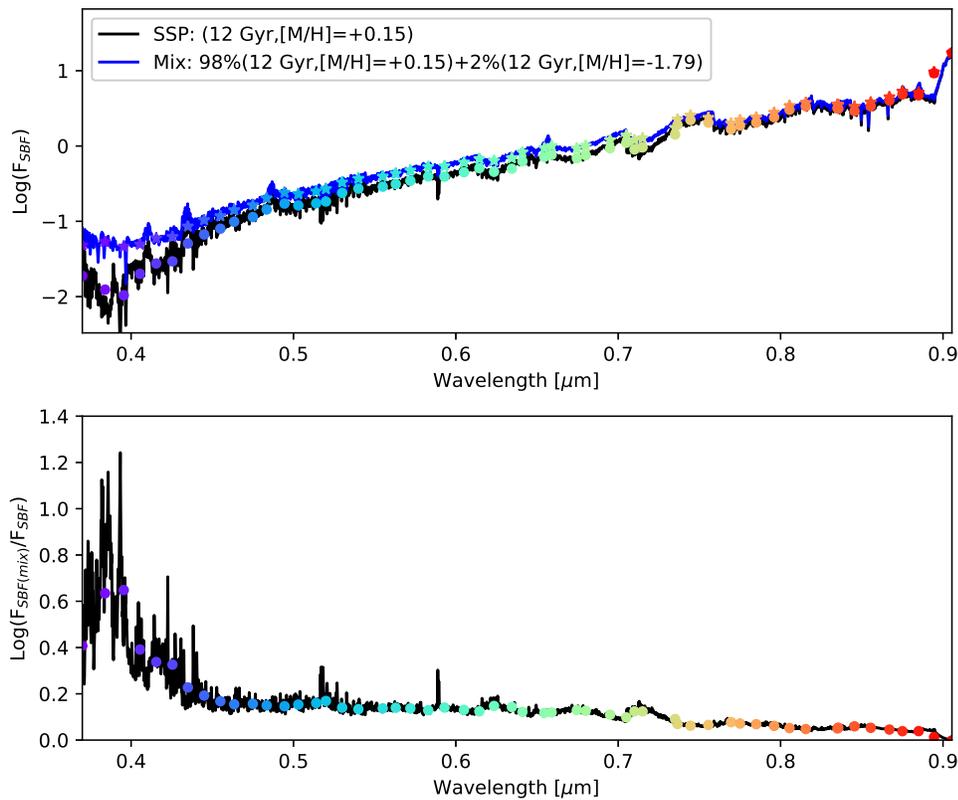} 
 \caption{The upper panel shows two SBF spectra, one corresponding to a metal-rich old SSP (as indicated in the inset) and another one where this SSP is combined with a $2\%$ (in mass fraction) of an equally old but very metal-poor SSP (see the inset). The spectroscopic fluctuation magnitudes corresponding to the narrow band filters of the J-PAS survey are indicated with circles of varying colours. The lower panel shows the resulting residuals between these two models (in magnitude). Note that such differences are significant and can be easily captured with the survey data.}
   \label{fig2}
\end{center}
\end{figure}

The potential of the SBFs for the stellar population studies can be further optimised with the aid of narrow band observations such as those of the J-PAS survey (Javalambre Physics of the Accelerating Universe Astrophysical Survey) (\cite[Cenarro et~al. 2010]{Cenarro10}), which is composed of $56$ narrow-band filters covering the optical spectral range. Such observations represent an intermediate step between imaging and spectroscopy, with the important advantage of achieving a high photometric precision ($\sim0.05$\,mag). Fig.~\ref{fig2} illustrates this potential: tiny contributions from very metal-poor old components lead to magnitude differences above $0.1$\,mag in the optical range and much larger blueward $\sim4500$\,\AA. Such differences can be easily detected with the survey data for nearby galaxies. Urge investigating and proposing diagnostic SBF diagrams based on these filters, which could also help us to uncover the distribution of these metal-poor contributions within nearby galaxies.

\end{document}